\documentclass{kapproc} % Computer Modern font calls

\usepackage{latexsym}
\usepackage{epsfig}
\usepackage{amsmath}
\usepackage{amsfonts}
\usepackage{amssymb}

\normallatexbib

\begin{document}

\articletitle[Aharonov-Bohm Effect in the Quantum Hall Regime and 
Laplacian Growth Problems]{Aharonov-Bohm Effect\\ 
in the Quantum Hall Regime\\ and fingering instability}

\author{P. Wiegmann} 
\affil{James Franck Institute and Enrico Fermi Institute\\
University of Chicago\\ 
5640 S.Ellis Avenue, Chicago, IL 60637, USA\\
and\\
Landau Institute for Theoretical Physics}
\email{wiegmann@uchicago.edu}

\chaptitlerunninghead{Quantum Hall effect and growth problems}

\begin{abstract}
The shape of an electronic droplet in
the quantum Hall effect is sensitive to  gradients of the magnetic
field,  even if they are placed outside the droplet.
Magnetic impurities cause a fingering instability
of the edge of the droplet, similar to the Saffman-Taylor fingering
instability of
an interface between two immiscible phases.
We discuss the fingering instability and some
algebraic aspects of the electronic states in a strong
nonuniform field.
%The notes are based on the recent paper \cite{ABWZ}.
\end{abstract}


\section{Aharonov-Bohm effect and the shape of electronic
droplets in a magnetic field}
\subsection{Introduction}
The Aharonov-Bohm effect is a striking manifestation of
interference in quantum processes. It has been observed in a number of
quantum mechanical and mesoscopic systems and proved to be an important
element of our  understanding of quantum physics.

In this notes we discuss yet another (so far just theoretical) realization of
the Aharonov-Bohm effect, now,  in a strong  magnetic field. The
discussion is based on the recent paper written in collaboration with O.
Agam, E. Bettelheim and A. Zabrodin \cite{ABWZ}.

Electrons confined in a plane in a strong
%quantizing
magnetic field form incompressible droplets
     trapped by an electrostatic  potential.  The area of the
droplet is quantized and is equal to $\pi N \ell^2$, where $N$ is a
number of electrons in the droplet and $\ell$ is a magnetic length. If
$N$ is large, the droplet is well described in a semiclassical manner.
It has a sharp edge distributed over a length $\ell$.

If magnetic field is uniform, the shape of the
droplet is determined by an equipotential line of the electrostatic
landscape. In the case of symmetric potential and a uniform magnetic field
the droplet is  a disk.

Let us now change magnetic field somewhere away from the droplet in a
manner that the magnetic field stays uniform in the area of the
droplet. For example we can do this by putting  some number of
Aharonov-Bohm fluxes or any sort of  magnetic impurities.
As electrostatic potential,  gradients of magnetic field  remove the
degeneracy of the Landau level and, therefore,
     affect the shape of the droplet.  However, the ways  electrostatic and
magnetic  forces shape the droplet are different.

Electrostatic potential affects the quantum droplet only if it is placed
inside the droplet. Its effect decays exponentially with the distance
from the droplet.
On the contrary, gradients of the magnetic field, even being placed
away from the droplet will  strongly affect the shape of the droplet.
Their effects decay slowly, as a power law in the distance from the
droplet.

Moreover, in the situation when  potential landscape is negligibly
flat, Aharonov-Bohm fluxes placed outside of the droplet cause a
fingering instability - an unstable  pattern of
fingers which grow with increasing the area of the droplet (Fig.~1).
A very similar
instability is known in non-equilibrium processes driven by diffusion
\cite{review}.

\begin{figure}
\begin{center}

\caption{A schematic illustration of the shape of  an
electronic droplet in a strong magnetic field when some additional
magnetic fluxes placed outside of the droplet. Electronic droplet is
stratified by semiclassical orbits. The area bounded by each orbit is
$\pi N\ell^2$}
\end{center}
\end{figure}

The effect of magnetic impurities is even more dramatic. Almost any
gradient of magnetic field at sufficiently large area of the droplet
    curves the edge so strongly  that segments with the curvature of the
order of inverse magnetic length  appear inevitably.
   At these segments the  semiclassical
description of the droplet and its edge states is no longer valid.

\subsection{Electronic droplet in the Quantum Hall regime.}

Consider first  $N$ spin-polarized electrons on a plane in a
uniform, perpendicular magnetic field $B_0>0$, in the
lowest Landau level:
\begin{equation}\label{1}
H=\frac{1}{2m}( -i\hbar \vec\nabla -\frac{e}{c}\vec A )^2.
\end{equation}
Degenerate states, written in the symmetric gauge,  have the form
$f(z)e^{-\frac{|z|^2}{2\ell^2}}$, where $f(z)$ is a holomorphic function.
Let us confine electrons in a flat symmetric potential well of large
radius $R$, well exceeding  $\ell\sqrt{N}$
   ($\ell=\sqrt{2\hbar c/eB_0}$ is a magnetic length).
The potential well lifts the degeneracy of the level such that a state
with higher angular  moment $n$ acquires  a higher energy. Near the
origin  the wave functions are close to the degenerate lowest Landau level
wave functions with given orbital momentum. Their orthogonal basis is:
\begin{equation}\label{2}
\psi_{n+1}^{(0)}=\frac{1}{\sqrt{\pi n!}}\frac{z^n}{\ell^{n+1}}
e^{-|z|^2/2\ell^2}.
\end{equation}
We say that $N$ particles form a droplet, when all first $N$
orbitals,
   $n=0,1\dots,N-1$  are occupied \cite{Laughlin}:
\begin{eqnarray}
\Psi^{(0)}(z_1,\cdots,z_N)&=&\mbox{det}\,
\psi_n^{(0)}(z_m)\big |_{n,m<N} \nonumber\\
&=&\frac{1}{\sqrt{N!\tau_N^{(0)}}}\Delta(z)
e^{-\frac{1}{2\ell^2}\sum_n |z_n|^2}.
\end{eqnarray}
Here
$\Delta(z)=\prod_{n<m\leq N}(z_n-z_m)=\mbox{det}\
(z_{m+1}^n)\big |_{0\leq n,m<N}$ is the Vandermonde
determinant and the
normalization factor
$(N!\tau_N^{(0)})^{-1/2}=\prod_{0\leq
n<N}{h_n^{(0)}}$
is the product of the normalization factors (\ref{2}) of
one-particle states \\ $h_n^{(0)}=(\sqrt{\pi n!}\ell^{n+1})^{-1}$.

In the semiclassical limit $N\gg 1$, this wave function describes a
circular shaped droplet of the radius $\ell\sqrt{N}$. In this limit all
arguments $z_n$ obey the saddle point equation
\begin{equation}\label{5}
\sum_{ m\neq n}^N\frac{2\ell^2}{z_n-z_m}=\bar z_n,
\end{equation}
and are uniformly distributed within a disk of the area $\pi N\ell^2$.
The wave function  decays exponentially if $z_n$ is found outside the
droplet.

Now consider the following arrangement (Fig.~1): the magnetic field remains
uniform in the area which includes the droplet (a disk with the radius
greater  than $\ell\sqrt{N}$). Away
from the droplet, the magnetic field is perturbed
$B(x,y)=B_0+\delta B$, in a manner that the nonuniform part does not
carry flux $\int \delta B dx dy=0$. Since
$\delta B=0$ in  the area of the droplet, the potential
$V(z)$  defined as  $\Delta V(z)=-\delta B/2B_0$ is harmonic.
The  gauge potential is deformed by a harmonic function\footnote{
We set $\hbar=e=c=1$ hereafter.}:
$A=A_y-iA_x=\frac{\bar z}{\ell^2}-2\frac{\partial}{\partial z}V(z)$.

Let us set the parameters $t_k$ to be
the  harmonic moments of the deformed magnetic field:
\begin{equation}\label{6} t_k=\frac{1}{\pi k }\int \frac{\delta
B(z)}{B_0}z^{-k} d^2z,
\end{equation}
For example, in the case of a few thin solenoids with fluxes
$ \Phi_a<\pi$ added at points $\zeta_a$, the harmonic moments are
$t_k=\frac{1}{2B_0}\sum_a\Phi_a\zeta_a^{-k}$.
The harmonic potential then is
\begin{equation}\nonumber
%\label{Vharm}
V(z)=Re \sum_{k\geq 1} t_k z^k.
\end{equation}

A nonuniform part in the magnetic field perturbs the wave function
   by a ``singular gauge transformation''
\begin{equation}\label{14}
\Psi(z_1,\cdots,z_N)
%\mbox{det}\,
%\psi_n^{(0)}(z_m)\big |_{n,m<N}
%\end{equation}
%\begin{equation}
=\frac{1}{\sqrt{N!\tau_N}}\Delta(z)
e^{-(\sum_n\frac{1}{2\ell^2} |z_n|^2-V(z_n))}.
\end{equation}
The saddle point equation (\ref{5}) is transformed  accordingly
\begin{equation}\label{51}
\sum_{ m\neq
n}^N\frac{2\ell^2}{z_n-z_m}={\bar z_n}-2\ell^2\frac{\partial}{\partial z}
V(z).
\end{equation}
This result holds in the limit when the radius of the confining
potential is very large. In this case the energy splitting of the lowest
Landau level due to the confining potential is less than the energy
splitting caused by gradients of the magnetic field.

The solution of this equations at large $N$ has been studied in Refs.
\cite{kkmwz}. The result is as follows:
all $z_n$ are uniformly distributed with the density $(\pi\ell^2)^{-1}$in
a domain  characterized by the following data,
\begin{itemize}
\item [-] the area of the domain is $\pi N\ell^2$;
\item [-] the harmonic moments of the exterior of the  domain
\begin{equation}\nonumber
t_k= -\frac{1}{\pi k} \int\!
{z^{-k}}{d^2z},\quad
k=1,\,2,\ldots
%\label{hrmonicmoments}
\end{equation}
(the integral runs over the exterior of the domain) are equal to the harmonic
moments of the nonuniform part of the magnetic field (\ref{6}).
\end{itemize}
If the boundary of the domain is smooth and single connected, these data
determine the domain.

We see that  gradients of magnetic field (say, Aharonov-Bohm fluxes)
placed away of the semiclassical orbits  affect the shape
of the droplet. The effect of the gradients dies slow with the distance
between the droplet and the position of the gradients. Indeed, if
  $L$ is a typical distance between the droplet and the gradients,
then $t_k\sim L^{2-k}$ decay slowly with $L$. In particular,  the
quadrupole moment
$t_2$ of the magnetic field is transferred to  a droplet from an arbitrary
distance. The third moment severely disturbs the shape of the droplet.
Its effect decays with the distance as $1/L$.

In the next paragraph we
argue that the distortion of the droplet  caused by a generic gradient
of the magnetic field  (magnetic impurities) not only strong, but
unstable.  The magnetic impurities cause  a fingering instability.
Afterwards, we discuss the origin of the Eq. (\ref{51}).

\subsection{Laplacian growth problem.}
Consider a process where the area of the droplet $\pi t=\pi N\ell^2$
grows, while the gradients of the magnetic field $\delta B$ remains
intact. This can be achieved by increasing the number of electrons
$N$ (by changing the gate voltage, for example), or by decreasing the
uniform part of the magnetic field.  In this process the moments $t_k$
are fixed. This leads to the following geometrical
problem:\begin{itemize}
\item [-]
   find the dynamics of a domain while its area
increases while harmonic moments $t_k$ remain fixed.
\end{itemize}
This problem has been discussed in the context of
pattern formations in non-equilibrium  processes
when a front between two immiscible phases advances  with the normal
velocity proportional to the gradient of a harmonic field -
a mechanism often referred as
Laplacian growth (for a review see, e.g.,  \cite{review}).

Viscous  or Saffman-Taylor fingering  is one of the most
studied  instabilities of this type. It  occurs at the interface between
two incompressible fluids with different viscosities when a less viscous
fluid is injected into a more viscous one in a 2D geometry
(typically, the fluids are confined in the Hele-Shaw  cell
-- a thin gap between two  parallel  plates - or in porous media
\cite{LG}).

In a thin cell, the local velocity of a viscous fluid
is proportional to the gradient of pressure:
$\vec v = -\vec\nabla p$. Incompressibility
implies that the pressure $p(z)$ is a harmonic function
of $z=x+iy$ with a sink at infinity:
\begin{equation} \label{D1}
\nabla^2 p(z)=0,\quad p(z)\to
-\frac{1}{2}\log|z|,\quad |z|\to \infty
\end{equation}
If the difference between viscosities is large,
the pressure is constant  in the less
viscous fluid and, if the surface tension is
ignored, it is  also constant (set to zero) on the interface.
Thus on  the interface
\begin{equation}
p(z)=0,\quad
        v_n =-\partial_n p(z). \label{D'Arcy}
\end{equation}
If less viscous liquid is supplied  through the origin with a constant
rate, the area $\pi t$  of the less viscous fluid grows linearly with time
$t$.

A simple consequence of the growth process defined by these equations
(\ref{D1},
\ref{D'Arcy}) is that harmonic moments of the viscous fluid domain,
\begin{equation}\nonumber
t_k= -\frac{1}{\pi k} \int\!
{z^{-k}}{d^2z},\quad
k=1,\,2,\ldots
%\label{hrmonicmoments}
\end{equation}
where the integral runs outside of the droplet, do not change in time
\cite{R}. They are initial data of evolution. Indeed,
$$\frac{ d}{dt}t_k
=\frac{1}{\pi k}\oint_{\mbox{interface}}
z^{-k}\partial_n p(z)|dz|
=0
$$
since the pressure  is a harmonic function and is a  constant on the interface.
Conservation of the harmonic moments is an
equivalent formulation of Laplacian growth, where surface
tension is ignored (\ref{D'Arcy}).

We conclude that the growth of the
semiclassical electronic droplet in a strong magnetic field is
equivalent to the propagation  of a "water" drop  (less viscous liquid)
in "oil" (more viscous fluid). This result is not surprising: both
dynamics are determined by the condition of incompressibility.

\subsection{ Fingering instability, finite-time singularities and
destruction of edge states.} The Saffman-Taylor problem
has been intensively studied  experimentally and analytically. It has
been found that a small (almost arbitrary)  deviation from a circular
form of the initial shape of the droplet is unstable.
The droplet forms a pattern of growing fingers whose shapes become
complex as the area of the droplet increases \cite{review}.
Infact, the situation is even more dramatic. It is known that  some
fingers  develop cusp-like singularities within a finite time of growth
\cite{singularities}, i.e., when the area of the droplet is finite.
In other words, fingers growing with $N\ell^2$, become thinner, and
reach the atomic/molecular scale at the finite area of the droplet.

Similarly, one can cause a fingering instability
of the quantum Hall droplet
by changing the
gradients of the magnetic field at fixed area. Fingers  will be driven
to a cusp-like singularity by adiabatically changing any of the harmonic
moments $t_k$.
When this occurs, the Laplacian growth equations (\ref{D1},\ref{D'Arcy})
are no longer valid. At this point corrections obtained from the
Navier-Stokes equations must be taken into account. They introduce a
microscale in the form of the surface tension,
that stops the curvature
of the interface.
Another mechanism to cure the singularities
is the discretization of the liquid. In this case one assumes that
the "water" domain consists of small particles
of non-vanishing size \cite{F}.

The quantum Hall effect  may be considered as a quantum version of the
Laplacian growth problem \cite{ABWZ}.
It also provides an attractive mechanism of regularizing
cusp-like singularities on the scale of the magnetic length,
as it will become apparent in the following.

For the electronic droplet, a singularity means that,
by increasing the number of particles at fixed gradients of magnetic
field, the curvature of some segments of  the droplet becomes so large
that the semiclassical description is no longer valid.
  The electronic states of a sharp segment of the edge of
the droplet are not separated from the bulk. They enjoy universal
conformal  properties that are very different from the conformal
properties of edge states on a smooth part of the edge.

It is
important  that a singularity occurs inevitably.
\subsection{Quantization of a singularity.}
Summing up,  at some point on a
quantum Hall plateau, the edge of the droplet becomes
very sharp and does not obey  the standard semiclassical
description. It  cannot be  described by a
conformal field theory. Edge states at the cusp-like  singularity of the
classical edge seem important in tunneling processes
and  in the transitions between plateaus.

Analysis of the singularity is equally important for Laplacian growth
problem. Quantum Hall effect provides a "quantized" version of the
Laplacian growth where no singularity is possible on a scale less than
magnetic length. Quantization  may be seen as yet another
regularization of singularity.
The study, which we do not present here, shows that  the
states at the singularity enjoy  universal scaling features,  depending
only on the qualitative character of the singularity. The
    analysis of the scaling behavior at the
singularity is technically involved. Its algebraic aspects are similar to
the  universal scaling behavior of random surfaces, intensively
studied in the context of 2D quantum gravity at $c<1$, the so-called
double scaling limit (for a review, see  \cite{DiFrancesco}).
  Physics of the
states on a sharp edge is a subject of current studies of the author.

In the rest of these notes, we review some algebraic aspects
of the dynamics of the electronic droplet in a nonuniform magnetic  field
and  its relation with the Laplacian growth.

\section{Algebraic aspects of electronic states in the quantum Hall regime
and Laplacian growth.}
\subsection{Laplacian growth as an evolution of conformal maps.}
The Laplacian growth can be conveniently reformulated  as a problem of
evolutions of  conformal maps.

Let $w(z,t)$ is a conformal map  of the exterior of the droplet  to the
exterior of the unit disk $|w|\geq 1$ in such a manner
that the source at $z=\infty$ is mapped to infinity.
In terms of the conformal map the pressure is
$p=-\frac{1}{2}\log| w(z,t)|$ and the
complex velocity in the viscous fluid is
$v(z)=v_x -i v_y =\frac{1}{2} \partial_z \log w(z)$.
On the interface, it is proportional to the harmonic measure:
\begin{equation}\label{v}
v_n(z,t) = \frac{1}{2}|w'(z,t)|.
\end{equation}
The complex velocity is conveniently written
        using the Schwarz function, $S(z)$:  this is an
analytic function in the domain containing  the contour
such that $S(z)=\bar z$ on the boundary \cite{Davis}.
The complex velocity is expressed in terms of this
function by $\partial_t S(z)$. The  equation (identity)
\begin{equation}\nonumber
%\label{Sch}
     \partial_t S(z)=\partial_z\log w(z).
\end{equation}
describes the evolution of the droplet under the condition that all
parameters $t_k$ are kept fixed.

One may be interested  in the evolution of the droplet under a change of
some particular $t_k$ if the area and all other moments are kept fixed.
This has been studied in Ref.\cite{kkmwz}. For references we list the
result here. The evolution  reads:
\begin{equation}\label{tkappa}
\partial_{t_k} S(z)=\partial_z H^{(k)}(z),\;\;\;\; k=1,2,\ldots
\end{equation}
were the $k$-th Hamiltonian is a nonnegative part of the $k$-th power
ofthe inverse conformal map $z(w)$. They are
$H_k=\Bigl (z^k(w)\Bigr )_{+} +\frac{1}{2}\Bigl (z^k(w) \Bigr )_{0}$.
The symbols $(f(w))_{\pm}$  mean the truncated
Laurent series where only  terms with positive (resp. negative)
powers of $w$ are kept, while $(f(w))_{0}$ is the
constant term ($w^0$) of the series. The
derivatives in the last equations  are taken at fixed $z$.

In refs.\cite{kkmwz}, the set of equations (\ref{tkappa}) was identified
with the dispersionless limit of the integrable Toda lattice hierarchy.
The compatibility of these equations  give a set of nonlinear equations
which describe the evolution of conformal maps under a deformation of the
domain. For example, the first equation of the hierarchy is written for
the conformal radius $r=\frac{1}{2\pi i}\oint\frac{dz}{w(z)}$ and
$u=-\frac{1}{2\pi i r}\oint{w(z)}{dz}$. They are dispersionless limit of 
Toda equation and Kadomtzev-Petviashvili (KP) equations:
\begin{equation}
\partial_{t_1}\partial_{\bar{ t_1}}\log r^2(t)=\partial_t^2 r^2(t)
\label{dT}
\end{equation}
\begin{equation}
3\partial_{t_2}^2u_n+\partial_{t_1}(
-4\partial_{t_3}u_n
+12u_n\partial_{t_1}u_n)=0.
\label{dKP}
\end{equation}
We will not develop this aspect further. See  refs.
\cite{kkmwz} for the details.

\subsection{The wave function in a nonuniform magnetic field.}
We now return to the problem of the electronic droplet in a nonuniform
magnetic field.
Since the magnetic field is uniform inside the droplet, the
one-particle wave functions in this area  are obtained by
linear combinations of the wave functions (\ref{2}) times the gauge factor
$e^{V(z)}$. They have the form
\begin{equation}\label{161}\psi_{n+1}(z)=P_n(z)e^{-\frac{|z|^2}{2\ell^2}+V(z)},
\end{equation}
where $P_n$ is a holomorphic polynomial of the degree $n$.

There are two equivalent ways for finding the polynomials.
One  uses the fact  that  the deformed  wave
functions are still orthogonal. Therefore the holomorphic polynomials are
bi-orthogonal with the measure $e^{-\frac{|z|^2}{\ell^2}+2V(z)}$. This
condition  uniquely determines the polynomials. Their explicit form is
known \cite{Mehta}. It is given by  a multiple integral
\begin{equation}\label{12}
P_n(z)=\kappa_n^{-1}\int
\Delta(\xi)\prod_{i\leq
n}(z-\xi_i)e^{-\frac{|z_i|^2}{2\ell^2}+V(z_i)}d^2\xi_i
\end{equation}
where the normalization factor $\kappa_n^2=n!(n+1)!\tau_n\tau_{n+1}$
and $\tau_n$ is the tau-function:
\begin{equation}\label{tau}
\tau_N= \frac{1}{N!}\int
|\Delta(\xi)|^2\prod_ne^{-\frac{|\xi_i|^2}{\ell^2}+2V(\xi_i)}d^2\xi_n\,.
\end{equation}
In the case of  a  uniform magnetic field the integrals  are computed
exactly:  $P_n^{(0)}(z)= \frac{1}{\sqrt{\pi
n!}}\frac{z^n}{\ell^{n+1}}$.

Computing the Slater determinant $\mbox{det}(\psi_{n+1}(z_m))$,
we obtain the multiparticle wave function
(\ref{14}) (we used the fact that
$\det P_n(z_m)=\frac{1}{\sqrt{\tau_{N}}}\Delta (z)$, where
$1/\sqrt{\tau_{N}} $ is the product of the
coefficients of the highest monomials of $P_n(z)$). The formula of the
tau- function (\ref{tau}) follows from the normalization condition for the
wave-function.

Another way is to obtaine the orthogonal set of one-particle states   as
an overlap between  $N+1$- and $N$-particle states (\ref{14}):
\begin{equation}\label{22}
%\label{33}
\psi_{N+1}(z)=\int\Psi(z,\xi_1,\dots,\xi_N)
\overline{\Psi(\xi_1,\dots,\xi_N)}\prod_{n\leq N}d^2\xi_n.
\end{equation}
This prompts the Eqs.(\ref{161},\ref{12}).

\subsection{Semiclassical states.}
At large $N$, one may treat the formulas (\ref{14},\ref{tau})
in the semiclassical approximation. This
immediately yields Eq.(\ref{5}), and to the shape of the droplet described
after this equation. It is interesting to go one step further to find
a semiclassical form of the wave function  in a nonuniform
magnetic field characterized by the harmonic moments  $t_k$ (\ref{6}).
     The result is sketched below (Ref.\cite{ABWZ}).

A semiclassical state is characterized by the orbit - a smooth, closed and
single connected loop  with the area $\pi n\ell^2$ and a given harmonic
moments
$t_k$. We recall that they are the moments of the nonunoform part of
the magnetic field (\ref{6}). The semiclassical form of the wave function
of this orbit (\ref{161}) is found to be
\begin{equation}\nonumber
\label{ps1}
\psi_N(z)\simeq
\Big({\frac{w'(z)}{2\pi\ell\sqrt{\pi}}}\Big)^{1/2}
e^{-\frac{1}{\ell^2}{\cal A}(z,\bar z)}e^{i\Phi(z)}
\end{equation}
Here $w(z)$ is a conformal map of the exterior of the orbit to th
exterior of the unit disk, a geometrical phase $2\pi\Phi(z)/\pi\ell^2$ is
the area of a sector bounded by a ray $\mbox{arg} z$ and some reference
axis. The   action
${\cal A}(z,\bar z)=\frac{1}{2}|z|^2-Re\,\Omega(z)$, where $\Omega(z)$ is
defined such that $\partial_z\Omega(z)=S(z)$ is the Schwarz function of
the domain. The action  is positive in the vicinity of the contour and
everywhere in the exterior domain. Its variation normal to the orbit reads
\begin{equation}\nonumber
%\label{2}
{\cal A}(z+\delta_n z)=|\delta_n z|^2-\frac{1}{3}\kappa(z) (\delta_n
z)^3+\ldots\,,
\end{equation}
where $\delta_n z$ is a normal deviation from  a point $z$ of the orbit
and $\kappa(z)$ is the curvature of the orbit.

A semiclassical state is localized at the
minimum of ${\cal A}(z)$, where the amplitude  has a sharp maximum.
All orbits have the same
harmonic moments $t_k$ and are differed by the area. The holonomy of the
state is $2\pi N$.

This result is easy to understand. The wave function
   (\ref{22},\ref{ps1}) is a matrix
element of the vertex operator at the edge of the quantum Hall state. The
edge states are conformal invariant. Therefore the vertex operators on the
edge of a  circular droplet (in a uniform magnetic field) and a  perturbed
droplet (in a nonuniform magnetic field) differ by the conformal
transformation: $\psi(z)\to (w'(z))^h\psi(z)$ where $h$ is a dimension of
the vertex operator.
In the integer Hall effect the edge excitations are free fermions:
$h=1/2$. Similar calculations for the semiclassical limit of
Laughlin's FQHE-$1/(m)$ states are expected to give
the prefactor in (\ref{ps1}) equal $(w'(z))^{1/2m}$.

In the classical approximation the amplitude of the wave function reads
\begin{equation}
|\psi_N|^2 \simeq\frac{1}{2\pi}|w'(z)|\delta(z),
\end{equation}
where the  $\delta$-function is localized on the orbit.

\subsection{Integrable structure of QHE states.}
An integrable structure for the dynamics of the semiclassical droplet
(evolution of conformal maps) suggests that the electronic states in
quantum Hall regime may also obey an integrable nonlinear equations. This
is, indeed, true.

Let us vary   magnetic field and follow an evolution of  the matrix
elements of electronic operators. They evolve according to the Toda
lattice hierarchy.
   A precursor of the integrability has been found in Refs.\cite{Cappelli}.
There, the operator content of  QHE was identified with
the $W_{+\infty}$ algebra.
We will address this issue in details elsewhere (see also
\cite{kkmwz,Z} and references therein). Below we will write the major
formulas.

The polynomials (\ref{12})  represent the  coherent states of
the   operator of magnetic translations
$Z=\ell^2(-2\partial_{ \bar z}+\bar A)$ in the arrangements where
nonuniform field is located outside of the droplet, i.e.,
when $ A-\ell^{-2}{\bar z}=-2\frac{\partial}{\partial z}V(z)$ is
a holomorphic function. This operator  annihilates all wave
functions  (\ref{161})  of the first Landau level $ Z\psi_n(z)=0 $
  and
acts as a multiplicator on the polynomials
$ZP_n(z)= zP_n(z)$.
The Hermitian conjugated
operator $\bar Z=\ell^2(2\partial_{ z}+ A)$ differentiates
the polynomials
$
\bar Z\psi_n=e^{-\frac{1}{\ell^2}|z|^2+V(z)}\ell^2\partial_zP_n(z).
$
In terms of these operators the Hamiltonian (\ref{1}) is
$H=\frac{1}{2m\ell^4}(Z\bar Z+\bar Z Z)$. Obviously
\begin{equation}\label{25}
[\bar Z,Z]=2\ell^2.
\end{equation}

As on  operator acting on polynomials, $Z$ is a lower triangular matrix
with the upper adjacent diagonal. The operator $\bar Z$ is a lower
diagonal matrix
\begin{equation}
Z_{nm}P_m(z)=zP_n(z),\quad Z_{nm}= 0\; \mbox{at}\;\; m>n+1.
\end{equation}
\begin{equation}
\bar Z_{nm}P_m(z)=2\ell^2\partial_zP_n(z),\quad \bar Z_{nm}= 0\;
\mbox{at}\;\; m\geq n.
\end{equation}
Their matrix elements depend on the magnetic field and are parametrized
by $t_k$.

The integrable hierarchy describes an evolution of the wave functions, or,
equivalently, the matrix elements of the operators $Z$ and $\bar Z$ as
functions of parameters $t_k$:
\begin{equation}\label{28}
\partial_{t_k}\psi_n(z)=H^{(k)}_{nm}(z)\psi_m(z)
\end{equation}
The commutative set of Hamiltonians $H^{(k)}_{nm}(z)$ are proved to be
a set matrices with zeros in  the lower triangular part. They are
\begin{equation}\label{75}
H^{(k)} =\bigl (Z^k \bigr )_{+}+\frac{1}{2}\bigl (Z^k \bigr )_{0}
\end{equation}
where $(Z^k \bigr )_{+}$ and $(Z^k \bigr )_{0}$ are upper triangular and
diagonal  parts of the $k$-th power of the matrix $Z_{nm}$.
In terms of operator $Z$ the evolution equations read
\begin{equation}
\label{73}
\frac{\partial Z}{\partial t_k} =
[H^{(k)}, Z],\quad \frac{\partial\bar Z}{\partial
t_k} =[H^{(k)}, \bar Z]
\end{equation}
These equations and the vanishing commutators among the Hamiltonians
give a set of nonlinear
equations for the matrix elements and coefficients of the polynomials.
For example, the equation for the ``quantum conformal radius''
   $r_n=Z_{n,n+1}$ and $u_n=\frac{Z_{n,n+2}}{Z_{n,n+1}}$ are the
celebrated Toda and KP equations:
\begin{equation}\label{173}
\ell^2\partial_{t_1\bar
t_1}^2\log
r^2_n=r^2_{n+1}-2r^2_{n}+r^2_{n-1},
\end{equation}
\begin{equation}
3\partial_{t_2}^2u_n+\partial_{t_1}(\ell^{-2}\partial_{t_1}^3u_n
-4\partial_{t_3}u_n
+12u_n\partial_{t_1}u_n)=0.
\label{KP}
\end{equation}
In the terminology of integrable hierarchies, the operators $Z$ and $\bar
Z$ are a pair of Lax operators; the wave function $\psi_n(z)$ is the
Baker-Akhiezer function; Eq. (\ref{25}) is the string equation; Eqs.
(\ref{28}-\ref{73})) are the Lax-Sato equation.
Finally, (\ref{tau}) represents the tau-function of the hierarchy.

The connection with the semiclassical description is transparent. The
classical limits ($\ell^2\to 0$) of the operators $Z$ and $\bar Z$
   are the coordinate of the droplet and its Schwarz function.
The classical (dispersionless) limit of the  Lax-Sato equations
describe the evolution of conformal maps (\ref{tkappa}). The Toda
equation (\ref{173}) is reduced to the dispersionless Toda equation
(\ref{dT}) for the conformal radius in the limit
$n\to\infty$ while  $t=n\ell^2$ is kept fixed.

\subsection{Random matrix representations.} 
The wave functions of the Quantum Hall effect
are naturally related to random matrices.
The square of the amplitude of the  multiparticle wave function
(\ref{14}) can be obtained as a result of the integration of
$e^{-\frac{1}{\ell^2}\mbox{tr}M M^\dagger+\mbox{tr}V(M,M^\dagger)}
$ over certain ensembles of complex matrices.  Here
$2V(M,M^\dagger)=\sum_k (t_kM^k+\bar t_k ( M^k)^\dagger)$.

One ensemble is $N\times N$ normal matrices
  with a given set of complex  distinct eigenvalues $z_1,\ldots,z_N$
\cite{Z,kkmwz}. We recall that the normal matrices are the complex
matrices with a relation
$[M,\,M^\dagger]=0$.  Integration over these matrices recovers
  (\ref{14}) up to a factor.

Another ensemble has been pointed to the author by M. Hastings. This is an
ensemble of arbitrary complex matrices \cite{G}. We recall this relation
briefly. Any complex matrix with distinct eigenvalues $z_1,\ldots,z_N$ can
be decmposed into as $M=U^\dagger(\mbox{diag}(z_1,\ldots,z_N)+R)U$, where
$U$ is a unitary matrix and $R$ is an upper triangular complex matrix.
Potential  $\mbox{tr}V(M, M^\dagger)=\sum_nV(z_n,\bar z_n)$ depends only
on eigenvalues, while  the measure of the integral $D[M]=D[U]
D[R]\, |\Delta(z)|^2$, and $\mbox{tr}M
M^\dagger=\sum_n|z_n|^2+\sum_{i>j}|R_{ij}|^2$ are factorized. The
volume of the unitary group $\int D[U]$ and the gaussian integration over
  matrix elements
$R_{ij}$ of the matrix $R$ contributes  just numerical factors. As a
result
$|\Psi(z_1,\ldots,z_N)|^2\sim \int DMDM^\dagger
e^{-\frac{1}{\ell^2}\mbox{tr}M M^\dagger+\mbox{tr}V(M,M^\dagger)}$.

Appearance of integrable hierarchies and random matrices
ties the Laplacian growth and the dynamics of quantum Hall edge states to
a number of important problems of theoretical and mathematical physics.

\begin{acknowledgments}
Useful dis\-cus\-sions with A. Caceres, A. Bo\-yar\-sky,  M. Hastings, L.
Le\-vi\-tov, M. Mi\-ne\-ev-\-Wein\-stein, A. Cappelli, V. Ka\-za\-kov, I.
Kos\-tov, L. Ka\-da\-noff,  O. Ru\-chay\-skiy, R. Teodorescu
and collaboration with O. Agam, E. Bettelheim and A. Zabrodin are
acknowledged. The work was supported by grants NSF DMR 9971332 and MRSEC
NSF DMR 9808595.
\end{acknowledgments}

\end{document}